\documentclass[journal]{IEEEtran}
\ifCLASSINFOpdf

\else
\fi

\usepackage{graphics}
\usepackage{epsf,graphicx}
\usepackage{latexsym,amssymb}
\usepackage{cite}
\usepackage{subfig}
\usepackage{url}
\usepackage{amsmath}
\usepackage{rotating}
\usepackage{pdflscape}
\usepackage{afterpage}
\usepackage{capt-of}

\usepackage{lipsum}
\begin{document}
\title{A Novel Chaos-based Light-weight Image Encryption Scheme for Multi-modal Hearing Aids}

\author{\IEEEauthorblockN{Awais Aziz Shah\IEEEauthorrefmark{1},
Ahsan Adeel\IEEEauthorrefmark{2}, Jawad Ahmad\IEEEauthorrefmark{1}, Ahmed Al-Dubai\IEEEauthorrefmark{1},
Mandar Gogate\IEEEauthorrefmark{1}, Abhijeet Bishnu\IEEEauthorrefmark{3}, \\Muhammad Diyan\IEEEauthorrefmark{1},  Tassadaq Hussain\IEEEauthorrefmark{1}, Kia Dashtipour\IEEEauthorrefmark{1}, Tharm Ratnarajah\IEEEauthorrefmark{3}}, and
Amir Hussain\IEEEauthorrefmark{1},\\
\IEEEauthorblockA{\IEEEauthorrefmark{1}School of Computing, Edinburgh Napier University, UK. \\ Email: \{a.shah, j.ahmad, A.Al-Dubai, mandar.gogate, m.diyan, t.hussain, k.dashtipour, a.hussain\}@napier.ac.uk}\\
\IEEEauthorblockA{\IEEEauthorrefmark{2}School of Mathematics and Computer Science, University of Wolverhampton, UK\\
Email: \{a.adeel@wlv.ac.uk\}}\\
\IEEEauthorblockA{\IEEEauthorrefmark{3}School of Engineering, University of Edinburgh, UK\\
Email: \{abishnu@exseed.ed.ac.uk, t.ratnarajah@ed.ac.uk\}}
}

\maketitle

\begin{abstract}
Multimodal hearing aids (HAs) aim to deliver more intelligible audio in noisy environments by contextually sensing and processing data in the form of not only audio but also visual information (e.g. lip reading). Machine learning techniques can play a pivotal role for the contextually processing of multimodal data. However, since the computational power of HA devices is low, therefore this data must be processed either on the edge or cloud which, in turn, poses privacy concerns for sensitive user data. Existing literature proposes several techniques for data encryption but their computational complexity is a major bottleneck to meet strict latency requirements for development of future multi-modal hearing aids. To overcome this problem, this paper proposes a novel real-time audio/visual data encryption scheme based on chaos-based encryption using the Tangent-Delay Ellipse Reflecting Cavity-Map System (TD-ERCS) map and Non-linear Chaotic (NCA) Algorithm. The results achieved against different security parameters, including Correlation Coefficient, Unified Averaged Changed Intensity (UACI), Key Sensitivity Analysis, Number of Changing Pixel Rate (NPCR),  Mean-Square Error (MSE), Peak Signal to Noise Ratio (PSNR), Entropy test, and Chi-test, indicate that the newly proposed scheme is more lightweight due to its lower execution time as compared to existing schemes and more secure due to increased key-space against modern brute-force attacks.
\end{abstract}

\section{Introduction}
With the recent advances in the hearing aids technology, artificial intelligence and machine learning is playing a vital role to make the technology more intelligent to overcome the traditional problem of surrounding environment noise. This can be achieved through introducing a combination of audio and visual technology to gather data both through visual information and audio to generate better audio output to the user. Here, machine learning plays a vital role in processing the data and producing a refined audio output. Since the complexity of the machine learning algorithms is high and the compact audio-visual hearing aids uses micro embedded hardware does have high computational processing power, therefore the data is transmitted from the audio-visual hearing aid towards the edge or cloud where the machine learning is performed. Here, a big challenge is the data privacy in terms of audio and visual data of the user. Traditional data encryption algorithms such as Advanced Encryption Standard (AES) and Rivest–Shamir–Adleman (RSA) are performing better in different domains but in the case of audio-visual hearing aids devices, there are several challenges such as the complexity of these algorithms is high and running them on the micro embedded hardware is an ambitious task, moreover if the computational complexity of the encryption is high, it will take more time in processing which is not feasible in the scenario of artificially intelligent hearing-aids where the delay must be very minimal. 

To overcome the aforementioned challenges, this paper builds on \cite{adeel2020novel} that previously proposed a light-weight chaos-based encryption scheme for audio-visual hearing-aids. The goal of this work is to reduce the processing time of the light-weight algorithm along with ensuring higher security and complexity. To achieve this goal a real-time audio/visual data encryption scheme based has been proposed based on Chaos-based Encryption using Tangent-Delay Ellipse Reflecting Cavity-Map System (TD-ERCS) map and Non-linear Chaotic (NCA) Algorithm, which was not previously used in \cite{adeel2020novel}. The novelty of this work is:
\begin{itemize}
\item The new chaotic maps used in this scheme are different as compared to the previous published work.
\item The scheme has been designed taking into account security, complexity, and execution time that would take less encryption time as compared to the previous one \cite{adeel2020novel} that would enable more security in the whole framework with very low execution time thus resulting in low latency.
\item One of the novelties of the paper is also the use of a state of the art Sbox \cite{hua2021design}. The paper proves the vulnerability of the previous Sboxes. 
\item The proposed scheme also uses increased keyspace which demonstrates higher security against different kinds of brute force attacks.
\end{itemize}

To demonstrate the effectiveness of the proposed real-time audio-visual encryption scheme for audio-visual hearing aids, comparisons have been made against different security parameters i.e., Correlation Coefficient, Key Sensitivity Analysis, Peak Signal to Noise Ratio (PSNR), Number of Changing Pixel Rate (NPCR), Unified Averaged Changed Intensity (UACI), Mean-Square Error (MSE), Entropy test, and Chi-test etc. which indicate that the newly proposed scheme is more lightweight due to its lower execution time as compared to existing schemes and more secure due to increased key-space against modern brute-force attacks.

The rest of the paper is organized as follows: Section 2 gives an overview of the chaotic maps used in the design of the encryption scheme, Section 3 describes the steps and working of the encryption scheme, the results obtained against different security parameters and comparison with existing methods are given in Section 4, and finally, Section 5 concludes the paper.

\section{Preliminaries} \label{Prel}
The proposed framework for the Audio-visual hearing aid is given in Figure \ref{framework}. The secure light-weight encryption developed in this work applies both confusion and diffusion process in addition to the substitution
process. The proposed scheme takes the plaintext image as an input and after applying the encryption steps listed in Section \ref{encrptionsteps} produces the encrypted image. This process includes applying chaotic maps TD-ERCS and NCA. The description of the used maps is given below:

\begin{figure*}[!t]
	\centering     
	\includegraphics[width=0.7\textwidth]{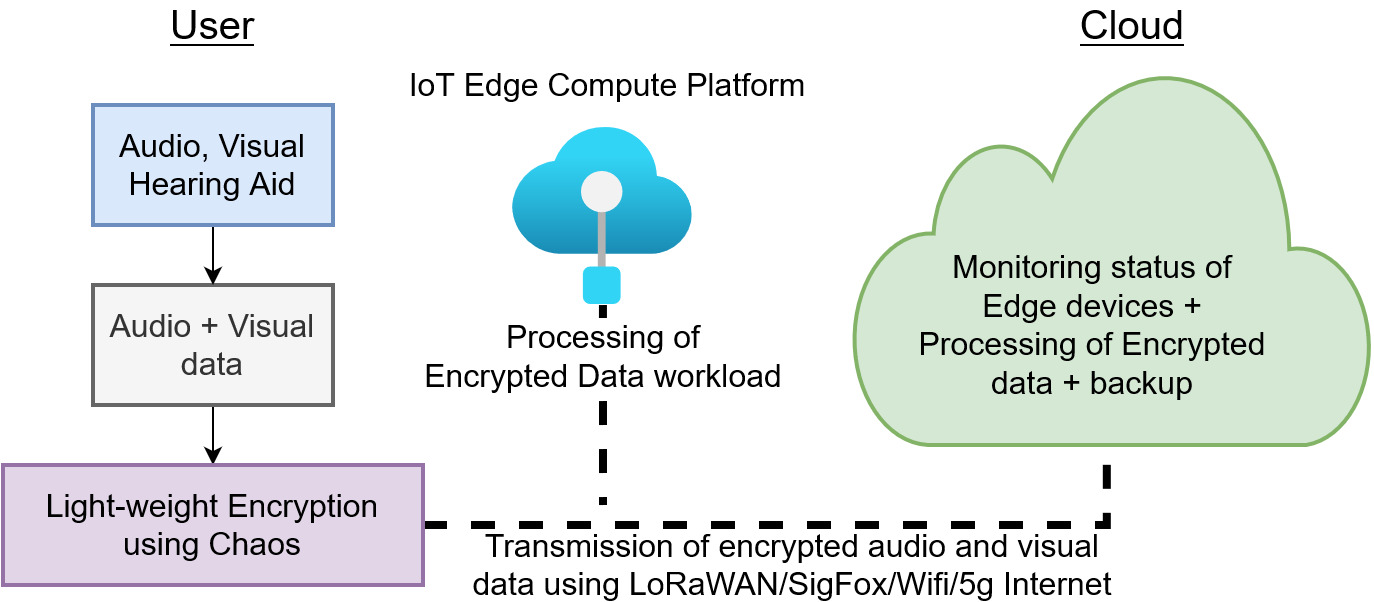} 
	\caption{Proposed 5G IoT-enabled Audio/Video hearing aid framework.}
	\label{framework}
\end{figure*}

 \subsection{Tangent-Delay Ellipse Reflecting Cavity-Map System (TD-ERCS)}
TD-ERCS chaotic map performs the row-wise and columnwise shufflig in the proposed scheme. It is a two-dimensional chaotic system and developed by \cite{li2005dynamical} based on a physical model of ellipse reflecting cavity. Numerous studies emphesize that this map satisfies many security features such as sensitivity dependence on initial conditions (SDIC), structure complexity, ergodicity, and deterministic pseudorandomness \cite{hussain2013novel, hussain2011analyses, mao2004novel} in addition to zero correlation in total field, positive Lyapunov exponent and equiprobability distribution \cite{behnia2007fast, gao2008new, rhouma2009ocml, patidar2009new, huang2009multi}. According to \cite{khan2017td} the TD-ERCS map can be described as:

\begin{equation} \label{eq1}
X_{i}=\frac{2 K_{i-1} Y_{i-1}+X_{i-1}\left(\mu^{2}-K_{i-1}^{2}\right)}{\mu^{2}+K_{i-1}^{2}}
\end{equation}

\begin{equation} \label{eq12}
Y_{n}=K_{i-1}\left(X_{i}-X_{i-1}\right)+Y_{i-1}, 
\end{equation}

\begin{equation} \label{eq13}
K_{i}=\frac{K_{i-j}^{\prime}-K_{i-1}+K_{i-1}\left(\left(K_{i-j}^{(\prime)}\right)^{2}\right)}{1+2 K_{i-j}^{\prime} K_{i-1}-K\left(K_{i-j}^{\prime}\right)^{2}}
\end{equation}

\begin{equation} \label{eq14}
K_{i-j}^{\prime}=\left\{\begin{array}{ll}
-\frac{X_{i-1}}{Y_{i-1}} \times \mu^{2} & \text { if } i<j \\
-\frac{X_{i-j}}{Y_{i-j}} \times \mu^{2} & \text { if } i \geq j
\end{array}\right.
\end{equation}
\begin{equation} \label{eq15}
Y_{0}=\mu \times \sqrt{1-X_{0}^{2}}
\end{equation}
$$
\begin{aligned}
K_{0}^{\prime} &=\frac{X_{0}}{Y_{0}} \times \mu^{2} \\
K_{0} &=-\frac{\tan \alpha+K_{0}^{\prime}}{1-K_{0}^{\prime} \tan \alpha}
\end{aligned}
$$

where
\begin{equation} \label{eq16}
\begin{array}{l}
\left\{\begin{array}{l}
X_{0} \in[-1,1] \\
\mu \in(0,1) \\
\alpha \in(0, \pi) \\
i=1,2,3, \ldots \\
j=2,3,4, \ldots
\end{array}\right.
\end{array}
\end{equation}

Here $\alpha$, $X_{0}$, j and $\mu$ are denoted as the seed parameters of TD-ERCS chaotic map.


where, $y_{n}$ are pseudo-random chaotic values,  $y_{n} \in (0,1)$, and $\lambda$ is the control parameter. Both $\lambda$ and $y_{0}$ serve as an initial condition and can called as key for chaotic pseudo-random number generation.

\subsection{Non-linear Chaotic (NCA) Algorithm}
 To further strengthen the encryption process, we apply the chaos-based NCA method \cite{choi2019color} to generate the encrypted image represented by $C_{i, j}$:
\begin{equation} 
C_{x, y}=N f_{2}^{x}\left(N f_{1}^{y}\left(S_{0}\right)\right)(1 \leq x, y \leq 512)
\end{equation}

\subsection{Secure Hash Algorithm (SHA)}
It uses a function from the plaintext message to generate the hash code. SHA has several variantions that aminly focus on the required size of the output such as SHA-1, SHA-256 and SHA-512 for the output 128 bits, 256 bits and 512 bits, respectively. Our proposed scheme uses SHA-512 such that $H(m)$ = $h$(512 bits). The suggested approach makes the Secret Key SHA-512 dependant, so that a tiny change in plaintext generates a completely new hash, and hence different initial key parameters.

\subsection{Affine Transformation}
The Sbox transformation has been used to apply another layer of security in the algorithm. It provides one to one mapping that is a unique plaintext symbol which is transformed to an other unique symbol. The following S-box of size 16 × 16 transformation is used in the proposed scheme \cite{hua2021design}:

\begin{table}[h]
\center
\caption{The Sbox used.}
	\setlength{\tabcolsep}{0.25em}
	\label{tab:assessment} 
\begin{tabular}{llllllllllllllll}
\hline
53  & 75  & 33  & 114 & 1   & 230 & 176 & 255 & 131 & 90  & 212 & 109 & 28  & 152 & 201 & 183 \\ 
130 & 17  & 80  & 172 & 256 & 47  & 10  & 147 & 85  & 237 & 105 & 126 & 180 & 203 & 214 & 56  \\ 
31  & 231 & 88  & 211 & 120 & 132 & 107 & 169 & 182 & 49  & 146 & 208 & 37  & 14  & 252 & 77  \\ 
193 & 12  & 164 & 32  & 52  & 119 & 185 & 136 & 219 & 102 & 45  & 79  & 250 & 82  & 238 & 149 \\ 
174 & 223 & 195 & 87  & 35  & 160 & 229 & 74  & 13  & 242 & 59  & 140 & 104 & 22  & 177 & 121 \\
228 & 58  & 189 & 249 & 205 & 21  & 158 & 210 & 44  & 71  & 175 & 8   & 134 & 112 & 115 & 81  \\ 
7   & 99  & 213 & 232 & 69  & 202 & 34  & 29  & 254 & 156 & 129 & 84  & 123 & 191 & 64  & 166 \\ 
248 & 117 & 98  & 43  & 18  & 221 & 76  & 190 & 151 & 196 & 96  & 233 & 161 & 51  & 138 & 15  \\ 
41  & 141 & 62  & 150 & 222 & 178 & 199 & 108 & 68  & 24  & 3   & 251 & 95  & 122 & 165 & 240 \\ 
125 & 198 & 159 & 142 & 239 & 241 & 83  & 48  & 170 & 5   & 184 & 50  & 215 & 73  & 27  & 100 \\ 
106 & 153 & 246 & 61  & 86  & 11  & 143 & 225 & 128 & 163 & 23  & 181 & 206 & 36  & 72  & 220 \\ 
224 & 244 & 9   & 186 & 137 & 168 & 54  & 91  & 97  & 127 & 78  & 19  & 157 & 236 & 39  & 194 \\ 
92  & 192 & 26  & 4   & 154 & 67  & 253 & 197 & 226 & 46  & 118 & 167 & 57  & 209 & 111 & 139 \\ 
70  & 94  & 135 & 207 & 103 & 60  & 216 & 116 & 25  & 187 & 245 & 145 & 227 & 173 & 2   & 42  \\ 
179 & 40  & 235 & 101 & 171 & 89  & 113 & 6   & 63  & 144 & 204 & 218 & 66  & 247 & 148 & 30  \\ 
155 & 162 & 124 & 65  & 188 & 110 & 20  & 55  & 200 & 217 & 234 & 38  & 16  & 133 & 93  & 243 \\ 
179 & 40  & 235 & 101 & 171 & 89  & 113 & 6   & 63  & 144 & 204 & 218 & 66  & 247 & 148 & 30  \\ 
155 & 162 & 124 & 65  & 188 & 110 & 20  & 55  & 200 & 217 & 234 & 38  & 16  & 133 & 93  & 243 \\ \hline
\end{tabular}
\end{table}

\section{Image Encryption Scheme} \label{encrptionsteps}
The detailed steps for the encryption process for the image using TD-ERCS, NCA, SHA-512, and Sbox of are listed below: 


\textbf{Step 1}: If the image is coloured, then conversion of image $I_c$ of size $A \times B$ to gray-scale will be done and the produced output $I_g$ and save resulted output to $\psi$. \\
\textbf{Step 2}: Apply hashing SHA-512 on gray-scale plaintext image $\psi$ and save hexadecimal hash value in variable $\theta$. \\
\textbf{Step 3}: Select first 12 hash values and save in $\kappa_1$ and select hash values from 13 to 24 and save value in $\kappa_2$.\\
\textbf{Step 4}: Convert hexadecimal values saved in $\kappa_1$ and $\alpha_1$ to decimal values and store result in $\kappa_1$, and $\beta_2$, respectively. \\
\textbf{Step 5}: Generate SHA-based initial conditions for TD-ERCS and NCA using below equations:
\begin{equation}
y_0 = \dfrac{\kappa_1}{2^{48}}
\end{equation}
\begin{equation}
x_0 = \dfrac{\kappa_2}{2^{48}}
\end{equation}
\textbf{Step 6}: Iterate TD-ERCS $A$ times and store chaotic values in $\alpha$. Randomly permute rows of gray-scale image $I_g$ using the sequence $\alpha$ and save values in $I_{rp}$.  \\
\textbf{Step 7}: Iterate NCA map $B$ times and store chaotic values in $\beta$. Randomly permute columns of $I_{rp}$ using the sequence $\alpha$ and save values in $I_{permuted}$.\\
\textbf{Step 8}: Apply Sbox and store the random values in $\gamma$.  \\
\textbf{Step 9}: Apply below operations on $\gamma$:
\begin{equation}
R_1 = \mbox{Mod}(\gamma \times 10^{14}, 256),
\end{equation}
\begin{equation}
R_2 = floor(R_1).
\end{equation}
\textbf{Step 10}: Rearrange row-vector $R_2$ in matrix form $R$ and Bit-wise XOR random matrix $R$ with $I_{permuted}$ to get $\phi$. \\
\textbf{Step 11}: Apply affine transformation on $phi$ and store values as a ciphertext image $C$. \\
For decryption, encryption steps are followed in reverse order.

\section{Encryption Results and Security Analyses}
\begin{figure}[!t]
	\centering     
	\includegraphics[width=0.2\textwidth]{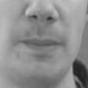} 
	\caption{Original Image.}
	\label{original_img}
\end{figure}

\begin{figure}[!t]
	\centering     
	\includegraphics[width=0.2\textwidth]{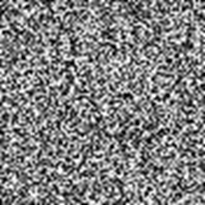} 
	\caption{Encryption Results.}
	\label{encrypted_img}
\end{figure}

\begin{figure}[!t]
	\centering     
	\includegraphics[width=0.5\textwidth]{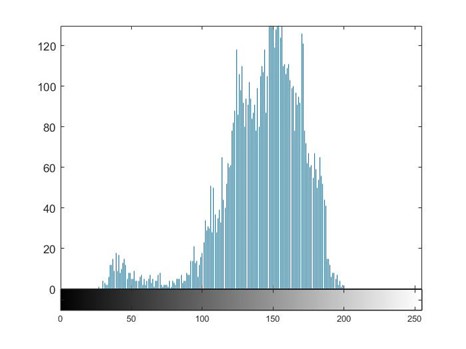} 
	\caption{Original Image Histogram.}
	\label{Histogram1}
\end{figure}

\begin{figure}[!t]
	\centering     
	\includegraphics[width=0.5\textwidth]{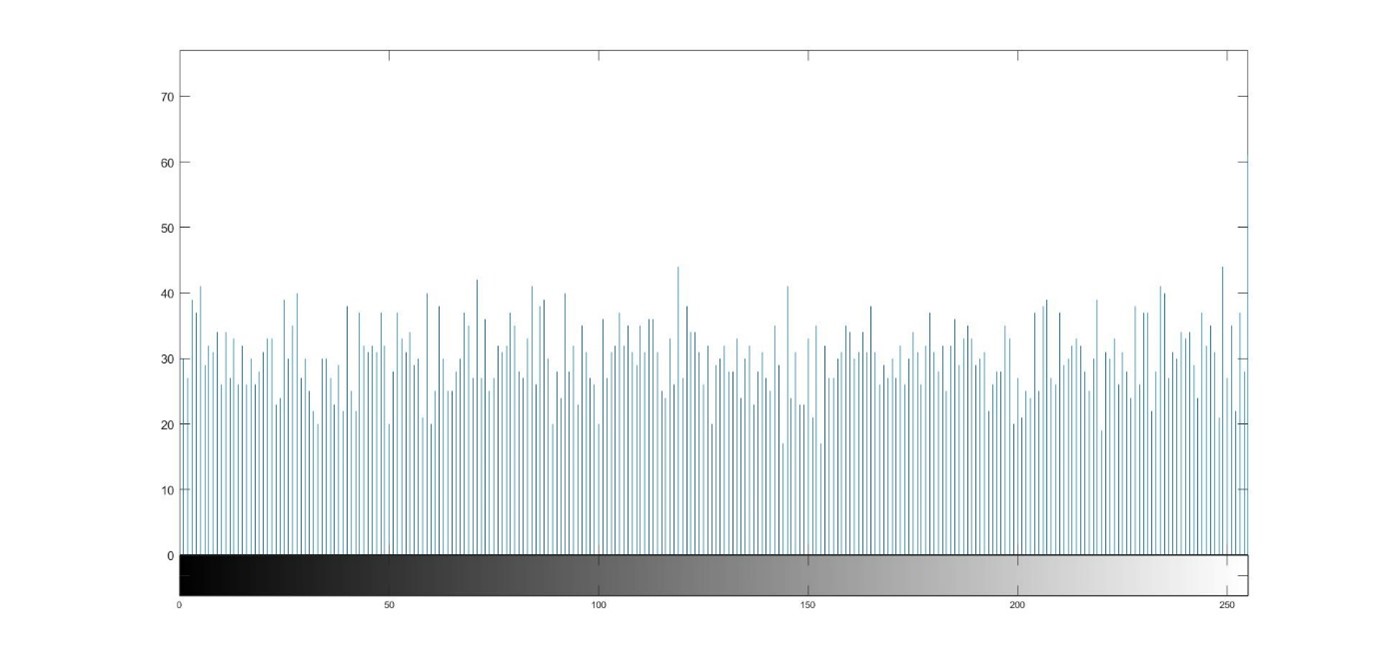} 
	\caption{Encrypted Image Histogram.}
	\label{Histogram2}
\end{figure}

To evaluate the performance of proposed scheme, an image with lips as seen in Fig. \ref{original_img}a has been selected. On the plaintext image, the proposed chaos-based encryption algorithm is used and the obtained results are shown in Fig. \ref{encrypted_img}b. The encryption results demonstrates that the proposed scheme completely conceal the plaintext information. To further analyze the results, histogram comparison is also been made in Fig. \ref{Histogram1} and Fig. \ref{Histogram2}, which confirms that the proposed scheme has an ideal histogram for encryption i.e., flat.

\begin{figure}[!t]
	\centering     
	\includegraphics[width=0.4\textwidth]{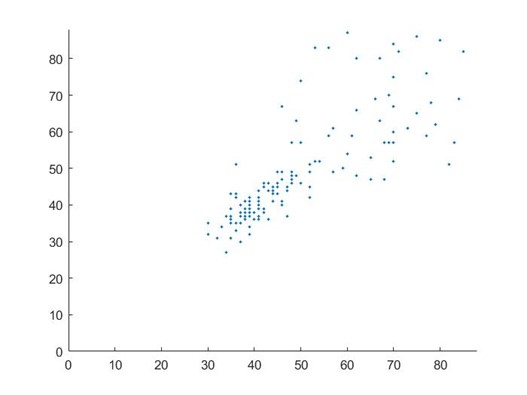} 
	\caption{Original Image correlation coefficient results.}
	\label{correlation1}
\end{figure}

\begin{figure}[!t]
	\centering     
	\includegraphics[width=0.4\textwidth]{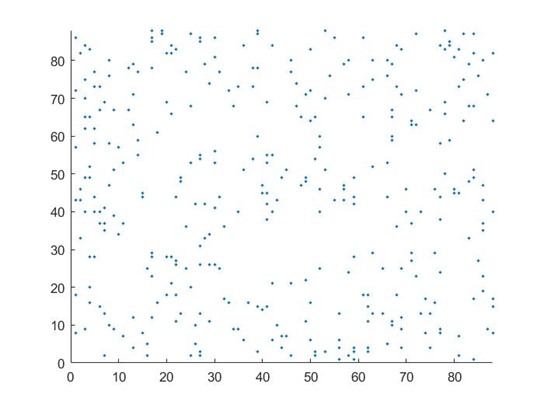} 
	\caption{Encrypted Image correlation coefficient results.}
	\label{correlation2}
\end{figure}

However, histogram alone can not justify the secureness of an algorithm, therefore several security analysis parameters are taken into account as previously defined in \cite{ahmad2017compression,khan2017novel,khan2017td}. The security metrics including correlation coefficient, entropy, contrast, energy, NPCR, and UACI are applied on the encryption scheme presented in this work.
Correlation is used as a metric to analyze the similarity between adjacent pixels. In a ideal situation, the correlation in all directions should be horizontal $(H_{CC})$, vertical$ (V_{CC})$ and diagonal $(D_{CC})$ directed closely towards zero. On the other hand, Entropy is an other important metric used to evaluate the resistance capability of the scheme against statical attacks. In a ideal situation, the entropy for a gray-scale encrypted image should be 8 bits. The higher values of contrast indicates higher quality of encrypted image. Sum of Squared Elements (SSE) in gray level co-occurrence matrix returns energy of an image. Low energy leve represents higher security of the image. In case of complete constant pixels, energy value is 1. Higher values of NPCR and UACI represents strong resistant against differential attacks. \cite{ahmad2017compression,khan2017novel,khan2017td}. \\
The plotting for correlation in vertical form is given in Fig. \ref{correlation1} and \ref{correlation2}. The observation highlights that the distribution of adjacent pixels in vertical direction is uncorrelated as compared to plaintext correlation. Mathematical values of correlations in vertical, horizontal and diagonal directions are outlined in Table \ref{tab:assessment}. From the tables, lower correlation values can be seen and hence proves the robustness of the proposed scheme. Moreover, other security  metrics are also shown in Table \ref{tab:assessment}. All metrics highlights the effectiveness and higher security of the proposed scheme. The required encryption/decryption time is less than 3 msec on a 8GB RAM, Ryzen 5 4600H CPU along with with 10/NVIDIA GTX 1650Ti 4GB GDDR6. Such a low processing time confirms that the proposed scheme is light-weight which is an effective solution for practical real-time applications.

\begin{table}[!t]
	\center
	\setlength{\tabcolsep}{0.4em}
	\renewcommand{\arraystretch}{1.8}
\caption{ Security assessment of the image.}
	\label{tab:assessment2} 
\begin{tabular}{lll}
\hline
Security Parameter         & Original Frame & Encrypted Frame \\ \hline
$D_{CC}$   & 0.9677         & 0.0033          \\
$H_{CC}$ & 0.9779         & -0.0018         \\
$V_{CC}$   & 0.9850         & -0.0018         \\
$Entropy$                    & 6.6875         & 7.9681          \\ 
$Contrast$                   & 0.1160         & 10.5347         \\ 
$Energy$                     & 0.2719         & 0.0158          \\ 
$Homogeneity$               & 0.9450         & 0.3889          \\ 
$NPCR$                       & N/A            & 98.5537         \\ 
$UACI$                       & N/A            & 32.8331         \\ 
$Key Sensitivity Difference$ & N/A            & 98.5537         \\ \hline
\end{tabular}
\end{table}

\section{Cross-comparison with existing techniques}
AES is an encryption scheme used widely for the privacy preservation of data in the industry, however, due to its complexity it poses challenge for deployment in constrained environments such as the Audio-visual hearing aids. This section performs cross-comparison of AES, the real-time chaos based encryption scheme presented in this paper, and the previously proposed on in \cite{adeel2020novel}. Table \ref{tab:videocomparison} presents the encryption time taken by the aforementioned algorithms against visual data and Table \ref{tab:audiocomparison}  highlights the time taken by encrypting audio data. It is evident from the results that chaos-based encryption schemes perform better in terms of processing time as compared to the traditional scheme AES, whereas the developed scheme takes less processing time as compared to the previous one in \cite{adeel2020novel}.

\begin{table}[!t]
\center
	\setlength{\tabcolsep}{0.4em}
	\renewcommand{\arraystretch}{1.8}
\caption{Comparison with existing methods for visual data.}

	\label{tab:videocomparison} 
	
\begin{tabular}{lllll} \hline
Method                             & \multicolumn{2}{l}{Single Image} & \multicolumn{2}{l}{75 Image frame} \\
                                   & Mean (Sec)    & Variance (Sec)   & Mean (Sec)     & Variance (Sec)    \\
AES                                & 0.290251      & 0.000662678      & 8.341109       & 0.001422          \\
\cite{adeel2020novel} & 0.00307       & 0.000003346      & 0.52068        & 0.003867          \\
This work                          & 0.00258       & 0.000002566      &
0.49189        & 0.003589         \\
\hline
\end{tabular}
\end{table}

\begin{table}[!t]
\center
	\setlength{\tabcolsep}{0.4em}
	\renewcommand{\arraystretch}{1.8}
\caption{Comparison with existing methods for Audio data.}

	\label{tab:audiocomparison} 

\begin{tabular}{lllll} 	\hline
Method                             & \multicolumn{2}{l}{Audio (512 frames)} & \multicolumn{2}{l}{Audio (38400 frames)} \\
                                   & Mean (Sec)       & Variance (Sec)      & Mean (Sec)        & Variance (Sec)       \\
AES                                & 0.058            & 0.00036             & 0.59437           & 0.00027              \\
\cite{adeel2020novel} & 0.014            & 0.01580             & 0.01580           & 0.02437              \\
This work                          & 0.012            & 0.00356             & 0.01466           & 0.00289             \\
\hline
\end{tabular}
\end{table}

\section{Conclusion}
This paper proposed real-time chaos-based lightweight encryption scheme for state of the art audio-visual hearing aids by enhancing the security of existing encryption schemes and making them more lightweight to work in low power embedded devices such as hearing aids. To accomplish this task several steps were performed including confusion and diffusion to encrypt the data, then shuffle the plaintext using two new chaos-based maps were used i.e., TD-ERCS and NCA. Logistic map diffusion has been performed using a newly proposed Sbox. The results obtained against standard security parameters such as Correlation Coefficient, , Peak Signal to Noise Ratio, Key Sensitivity Analysis, Number of Changing Pixel Rate, Mean-Square Error ,Unified Averaged Changed Intensity, Chi-test and Entropy test highlight the effectiveness of the proposed approach.
The future work of this paper will focus on evaluating the combination of audio-visual data in real-time scenarios.

\section{Acknowledgement}
This work is funded by the EPSRC COG-MHEAR programme grant (Grant no. EP/T021063/1)

\bibliographystyle{ieeetr}
\bibliography{ref}  

\end{document}